# Wave breaking onset of two-dimensional wave groups in uniform intermediate depth water

Arvin Saket[1], William L. Peirson[1], Michael L. Banner[2] and Michael J. Allis[1,3]

[1]Water Research Laboratory, School of Civil and Environmental Engineering, UNSW Sydney, King St., Manly Vale NSW 2093, Australia

[2]School of Mathematics and Statistics, UNSW Sydney, NSW 2052, Australia

[3]National Institute of Water and Atmospheric Research, Hamilton 3251, New Zealand

**Abstract**

Using the same measurement techniques as those of Saket *et al.* (2017), we have investigated the breaking threshold proposed by Barthelemy *et al.* (arXiv:1508.06002v1, 2015b) but for different classes of unforced unidirectional wave groups in intermediate water depths in the laboratory. The threshold parameter $B_x = U_s/C$ (where $U_s$ is the horizontal surface water particle velocity at the wave crest and $C$ is the wave crest point speed) which distinguishes breaking from non-breaking waves was found to be $0.835 \pm 0.005$ with the experimental uncertainty of each data point of $\pm 0.020$. This threshold is applicable to water depth to wave length ratios as low as 0.2 including the deep water conditions investigated by Saket *et al.* (2017). No dependence on peak spectral wavenumber was found at the scales achievable in a large-scale laboratory.

The present study provides more robust and universal characterisation of breaking in transitional water than the empirical non-dimensionalisation of Nelson (1994). The limiting wave height to water depth ratios of marginally breaking deep and intermediate water waves remain within 10 % of Nelson's values. However, it is shown that the effect of wave grouping can produce waves in shallower water that are at least 30 % greater in height than the limit proposed by Nelson (1994). The present study supports use of limits based on McCowan (1894) and Miche (1944) for coastal engineering design for marginal breaking waves. Three-dimensional and more strongly breaking waves in shallower water may yield wave heights higher than those measured during this study.

**Key words:** surface gravity waves, waves/free-surface flows, intermediate water wave breaking

## 1. Introduction

Waves at the transition to breaking are the critical design condition for marine and coastal structures (Silvester, 1974, *p*. 379*ff*). The quest to determine a reliable means of determining the onset of wave breaking has spanned 135 years since Stokes (1880) developed the first theoretical prediction of wave breaking. Over the past half century, many criteria have been proposed to determine the onset of wave breaking in intermediate water (wavelength to depth ratios greater than 0.5) with major contributions by Iversen (1952), Galvin (1969), Goda (1970), Weggel (1972) and Nelson (1994).

Based on local wave properties, theoretical criteria can be segregated into three categories: kinematic, dynamic and geometric criteria (Wu & Nepf 2002). Although the kinematic breaking criterion or geometric wave properties have been traditionally used as indicators of breaking onset in deep or shallow water, these criteria have largely failed and are not universally applicable (Melville 1994; Banner & Peirson 2007; Perlin, Choi & Tian 2013). Dynamic criteria based on energy flux rate show more promising ability to characterise the onset of wave breaking (Song & Banner 2002; Banner & Peirson 2007; Tian, Perlin & Choi 2008; Perlin *et al.* 2013; Derakhti & Kirby 2016).

Barthelemy *et al.* (2015b) proposed a breaking onset threshold based on a local threshold of wave energy flux in the crest region of a steep wave using numerical simulations of fully-nonlinear 2D and 3D wave packets in deep and intermediate water. For unforced water surfaces, this threshold can be defined as $B_x = U_s/C$ and maximises at the surface. Here, $U_s$ is the horizontal water surface particle speed in the wave propagation direction ($x$) and $C$ is the wave crest point speed. Once the ratio $B_x$ at the wave crest point (used interchangeably with the 'crest maximum elevation' throughout this paper) exceeds a critical threshold, the wave will progress to a breaking state. Barthelemy *et al.*'s threshold was found to be robust for long- and short-crested waves from deep to intermediate water.

Saket *et al.* (2017) investigated the breaking onset threshold of Barthelemy *et al.* (2015b) in the laboratory for different classes of two-dimensional unforced and wind-forced wave groups in deep water. Using a state-of-the-art thermal image velocimetry (TIV) technique to measure the velocity in the surface skin, they found that the threshold for the onset of breaking was 0.840 with an experimental uncertainty of ±0.016 and robust for different classes of waves in the absence and presence of wind.

Seiffert & Ducrozet (2016) examined numerically the breaking parameter proposed by Barthelemy *et al*. (2015b) for modulated, chirped and random waves in deep water. They found that the onset of breaking threshold to be robust for different types of wave groups with a value between 0.84 and 0.86.

The so-called breaker index $(H/d)_b$ (CERC, 1984) is fundamental to coastal engineering design methods and has been extensively investigated for waves propagating over horizontal beds. Here, $H$ is the wave height, $d$ is the mean water depth and the subscript $b$ denotes the value at breaking. Theoretical calculation by McCowan (1894) showed that the maximum value of $H/d$ prior to breaking of solitary wave in shallow water is 0.78 which is commonly used in coastal engineering design for horizontal and very gentle bed slopes. Miche (1944) theoretically showed that the limiting wave steepness at breaking $(H/\lambda)_b$ in intermediate water depth could be approximated using:

$$\frac{H_b}{\lambda_b} = 0.142 \ \tanh\left(\frac{2\pi d_b}{\lambda_b}\right), \qquad (1.1)$$

where $\lambda$ is wavelength and beyond this limit wave breaking will occur. Yamada *et al*. (1968) revised the value determined by McCowan (1894) to be 0.8261 (Goda 2010). More recent theoretical investigations for steady waves are presented in Longuet-Higgins & Fox (1996). Weggel (1972) used monochromatic waves to investigate experimentally the wave breaking as a function of wave steepness and submerged beach slopes and developed a suite of design curves. Experimentally, he found a breaker index of 0.78 in transitional water of constant depth. Further major laboratory studies and field observations on steeper slope showed that in shallow water the breaker index depends on the steepness and beach slope (Iversen 1952; Galvin 1969; Goda 1970; Weggel 1972; Nelson 1994).

Extensive work in the laboratory and in the field by Nelson (1985, 1994) found that the maximum wave height to water depth ratio for shallow water waves propagating over a horizontal bed could not exceed the value of 0.55. The differences between the findings of Weggel (1972) and Nelson (1994) are perplexing and remain controversial. Based on his experimental results, Nelson (1994) proposed a relationship to predict the limiting wave heights using a non-dimensional parameter $F_c$ as follows:

$$(H/d)_{max} = \frac{F_c}{22 + 1.82 F_c} \qquad (1.2)$$

$F_c$ was adopted from Swart & Loubser (1979) and is defined as:

$$F_c = \frac{g^{1.25} H^{0.5} T^{2.5}}{d^{1.75}} \qquad (1.3)$$

where $g$ is the acceleration due to gravity (m s$^{-2}$) and $T$ is the wave period (s). Riedel & Byrne (1986) carried out a laboratory investigation using monochromatic and random trains over a horizontal bed flume. Their experimental study showed that the values of $(H/d)_b$ ranged from 0.44 to 0.54 with $F_c$ between 49 and 150 for random waves and the largest value was 0.54 with $F_c$=150. They concluded that the limiting ratio of 0.55 proposed by Nelson (1985) applies equally well to random waves. Note that their experiments were conducted in the intermediate water depth and their measured value of $H/d$ actually exceeded Nelson's curve by up to 15 %. Gourlay (1994) measured the transformation of regular waves over a laboratory model of a coral reef. It was found that the largest wave height to water depth ratio never exceeds 0.55 for shallow water waves and the limiting wave height increases by increasing $F_c$ or the bed slope. Massel (1996) evaluated the maximum possible wave height limit over a horizontal bed theoretically in intermediate water depths using monochromatic waves. The maximum wave height was found to be less than $0.6d$ which was seen as being in agreement with the limit proposed by Nelson (1994).

Dack & Peirson (2005) investigated whether the spatial development of wave groups can influence the breaker index. They carried out a laboratory experiment to investigate the breaker indices of uni-directional bimodal group waves propagating above a horizontal bed. They found a maximum breaking wave height to water depth ratio of exceeding Nelson's curve by 22 %. They concluded that the intra-wave group interactions can play a key role in determining the initiation of wave breaking in intermediate water. Recently, Barthelemy *et al.* (2013) extended numerically this work using focussed wave groups propagating over a flat bed and found the value of $(H/d)_{max} = 0.605$.

Given the developing support for the dynamically-based threshold for the onset of breaking proposed by Barthelemy *et al.* (2015b), this present laboratory investigation has systematically investigated the onset of breaking of water wave groups propagating in intermediate and constant depth. The impact of wave groups on the breaker index is examined using both spatially-focussed and bimodal wave groups propagating over a horizontal bed.

## 2. Methodology

The experiments were carried out in the two-dimensional wave tank that was 30 m long, 0.6 m wide and 0.6 m deep with glass sidewalls and a solid floor located at UNSW Sydney, Water Research Laboratory (WRL) in Manly Vale, Australia. The tank configuration was identical to that used by Saket *et al.* (2017), except that the flexible wave paddle was replaced with a piston paddle system for intermediate depth conditions. Three water depths $d = 0.35$ m, $d = 0.22$ m and $d = 0.19$ m were selected to generate intermediate water waves. The change in depth created the possibility of reflections from a sluice gate behind the paddle. To minimise the reflections from the sluice gate behind the paddle, flexible reticulated polyester-urethane foam was positioned between the paddle and the gate (figure 1). The tunnel roof height was reduced to 0.12 m for optical reasons and used as the support of the measurement apparatus. No wind was used during the present study. The fan and the honeycomb flow guide used by Saket *et al.* (2017) were removed from the tunnel. Otherwise, all configurations, dimensions and distances were identical to those of Saket *et al.* (2017).

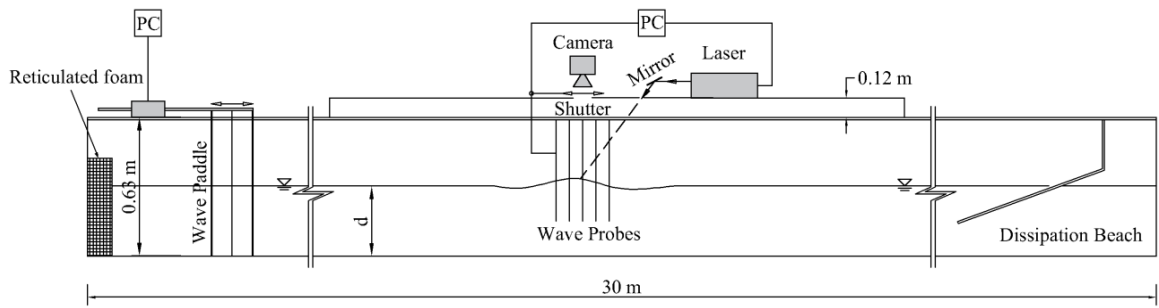

Figure 1. Schematic of the experimental set-up (not to scale). The water depth $d$ was 0.35 m, 0.22 m and 0.19 m to generate intermediate water waves. Further details can be found in Saket *et al.* (2017).

The thermal image velocimetry technique developed by Saket *et al.* (2016) was used to measure the horizontal water particle velocities $U_s$ at the crests of waves transitioning through the group maximum. The entire TIV system consisted of a thermal imaging camera, an irradiating source, a computer-controlled shutter, and a computer controlling the system components. Its operation is described in detail in Saket *et al.* (2016) and its validation for monochromatic waves is presented in Saket *et al.* (2017). Its configuration and all accompanying wave probe measurements to determine crest point speeds for this present study were identical to those used by Saket *et al.* (2017).

The wave generation signals used during this present study were selected to be identical to three deep water cases used by Saket *et al.* (2017).

The sensitivity of the breaking threshold was also investigated in relation to wave scale. The shallowest Class 2 three wave case with the water depth of $d = 0.19$ m was scaled using Froude scaling to a water depth of 0.26 m to determine the impact of changing wave length on the breaking threshold.

Each experiment commenced with the entire measurement apparatus being moved to the far end of the flume and the tank water surface cleaned as described in Saket *et al.* (2017). By systematically increasing the paddle amplitude, the forcing required and location of maximum recurrence and marginal breaking were determined. As described in Saket *et al* (2017) the maximum recurrent or maximum non-breaking threshold was defined for the condition which there was no residual disturbance of the water surface observed as waves transitioned through the group envelope maximum. Further slight but sufficient augmentation of paddle forcing triggered a significant change in dominant wave behaviour (marginal breaking) at the initial break point. In each case, consistent breaking was observed to commence at the point of maximum wave group amplitude and evidenced by a bulge in the forward face of the wave which was accompanied by trailing ripple disturbances of the surface. The resulting paddle amplitudes ($A_p$) are presented in table 1 for all conditions tested.

The entire measurement apparatus was then positioned above the location of the group maximum for each case and measurements of local wave characteristics and surface current were conducted. The elevation of the wave group maximum was defined as the local wave amplitude (*a*) above the mean water level. The horizontal distance between the adjacent mean water level locations spanning the group maximum determined the local crest half-wavelength ($\lambda_c = \lambda/2$) and the crest steepness $S_c = a\,\pi\,\lambda_c^{-1}$. The wave height *H* was determined as the vertical distance between the group maximum elevation and the average of the elevation of the troughs immediately preceding and following, observed at the same location. The linear wavelength ($\lambda_0$) was determined from the peak frequency (or the mean frequency in the case of Class 2 waves).

## 3. Results and discussion

3.1 Breaking Threshold

The average measured crest point speeds and horizontal surface water velocities at the location of the group maximum are summarised in table 1. The values obtained from each data set are shown within the table 1. Uncertainty expressed as the standard error of between 6 and 8 individual measurements are presented at the top of each column. The measured crest steepness $S_c$ results showed that the local steepness at breaking systematically increased from deep to intermediate water. Note that the local steepness levels between the marginal breaking Class 2 and maximum recurrent Class 3 waves overlap substantially.

The results of the measured velocities indicated that, in contrast to the crest surface water velocity, the changes of the crest point speed for all wave groups from maximum recurrence to marginal breaking are small. Therefore, the crest water surface velocity plays the dominant role in determining the onset of breaking in intermediate water as in deep water (Saket *et al.* 2017). The difference between crest surface velocities of marginal breaking and maximum recurrence to the crest point velocity ratio for all cases from deep to intermediate water is illustrated in figure 2. Although there is no systematic change for the Class 3 waves, the Class 2 cases show a systematic increase in water surface velocity across the transition from non-breaking to marginally breaking as the relative depth decreases.

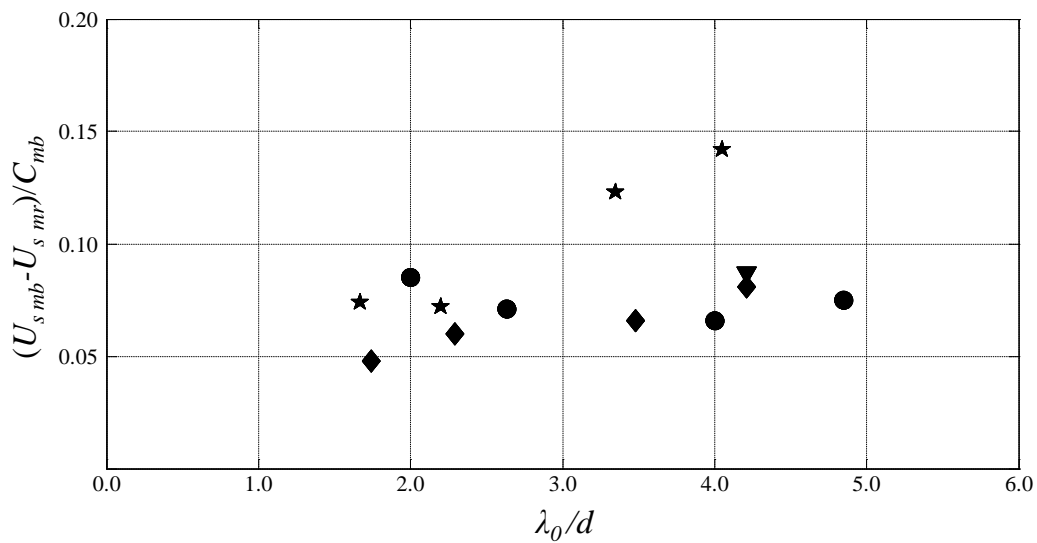

Figure 2. The deep water wavelength to water depth ratio versus the change in crest surface water velocities from maximum recurrence to marginal breaking divided by the marginal breaker crest point speeds for C2N3 (♦), scaled C2N3 (▼), C2N5 (★) and C3N9 (●) wave packets from deep to shallow water.

Table 1 - The ratio of linear deep water wavelength and water depth $\lambda_0/d$, wave paddle amplitude ($A_p$), the average and standard deviation of measured crest steepness ($S_c$), crest point speed ($C$), crest point water surface particle velocity ($U_s$), ratio of the velocities at the crest point $U_s/C$, non-dimensional parameter $F_c$, wave height to water depth ratio $H/d$ and non-dimensional parameter $\lambda \rho^{0.5} g^{0.5} \sigma^{-0.5}$ for maximum recurrence and marginal breaking waves.

| Case | | $\lambda_0/d$ | $A_p$ | $S_c$ | $C$ (m s$^{-1}$) | $U_s$ (m s$^{-1}$) | $U_s/C$ | $F_c$ | $H/d$ | $\lambda\sqrt{\rho g/\sigma}$ | $A_p$ | $S_c$ | $C$ (m s$^{-1}$) | $U_s$ (m s$^{-1}$) | $U_s/C$ | $F_c$ | $H/d$ | $\lambda\sqrt{\rho g/\sigma}$ |
|---|---|---|---|---|---|---|---|---|---|---|---|---|---|---|---|---|---|---|
| | | | (×10$^{-2}$ m) ±0.015 | ±0.015 | ±0.015 | ±0.021 | ±0.022 | — | — | — | (×10$^{-2}$ m) ±0.018 | ±0.018 | ±0.014 | ±0.027 | ±0.021 | — | — | — |
| C2N3 | | 1.74 | 1.032 | 0.402 | 0.959 | 0.791 | 0.825 | 5.447 | 0.150 | 245 | 1.067 | 0.422 | 0.964 | 0.837 | 0.868 | 5.703 | 0.158 | 248 |
| | | 2.29 | 1.061 | 0.431 | 0.958 | 0.768 | 0.802 | 9.092 | 0.201 | 239 | 1.103 | 0.450 | 0.967 | 0.826 | 0.855 | 9.254 | 0.216 | 238 |
| | | 3.48 | 1.175 | 0.440 | 0.887 | 0.726 | 0.819 | 20.866 | 0.303 | 225 | 1.326 | 0.470 | 0.904 | 0.786 | 0.869 | 22.086 | 0.316 | 226 |
| | | 4.21 | 1.417 | 0.459 | 0.870 | 0.688 | 0.790 | 29.706 | 0.345 | 218 | 1.504 | 0.491 | 0.896 | 0.761 | 0.849 | 31.062 | 0.375 | 216 |
| Scaled | | 4.21 | 2.049 | 0.464 | 1.028 | 0.794 | 0.772 | 29.279 | 0.362 | 300 | 2.094 | 0.495 | 1.055 | 0.886 | 0.839 | 30.844 | 0.373 | 301 |
| C2N5 | | 1.67 | 0.856 | 0.431 | 0.944 | 0.753 | 0.797 | 5.218 | 0.149 | 228 | 0.863 | 0.451 | 0.963 | 0.824 | 0.855 | 5.441 | 0.155 | 230 |
| | | 2.20 | 0.885 | 0.435 | 0.950 | 0.743 | 0.782 | 8.146 | 0.190 | 226 | 0.913 | 0.468 | 0.965 | 0.812 | 0.842 | 8.971 | 0.202 | 227 |
| | | 3.35 | 1.112 | 0.446 | 0.893 | 0.693 | 0.776 | 19.283 | 0.293 | 218 | 1.133 | 0.486 | 0.921 | 0.806 | 0.875 | 20.023 | 0.310 | 217 |
| | | 4.05 | 1.525 | 0.451 | 0.867 | 0.654 | 0.783 | 28.759 | 0.352 | 213 | 1.564 | 0.499 | 0.903 | 0.782 | 0.866 | 30.076 | 0.378 | 214 |
| C3N9 | | 2.00 | 1.733 | 0.478 | 1.191 | 0.936 | 0.786 | 13.300 | 0.261 | 389 | 1.756 | 0.486 | 1.240 | 1.041 | 0.840 | 13.559 | 0.262 | 390 |
| | | 2.63 | 1.762 | 0.482 | 1.209 | 0.955 | 0.790 | 21.084 | 0.322 | 354 | 1.815 | 0.520 | 1.232 | 1.043 | 0.846 | 21.833 | 0.344 | 360 |
| | | 4.00 | 2.004 | 0.533 | 1.137 | 0.918 | 0.808 | 48.210 | 0.454 | 297 | 2.035 | 0.560 | 1.170 | 0.995 | 0.851 | 49.572 | 0.475 | 298 |
| | | 4.85 | 2.269 | 0.558 | 1.109 | 0.879 | 0.792 | 67.481 | 0.502 | 266 | 2.274 | 0.574 | 1.123 | 0.963 | 0.854 | 69.188 | 0.515 | 267 |

Table spans two parts: "Maximum recurrence" (left) and "Marginal breaking" (right).

The value of energy flux ratio $B_x = U_s/C$ was determined for each case and has been presented as a function of the local crest steepness in figure 3. The threshold for the breaking onset is robust for different types of wave packets in spite of the changing water depth. With this expanded data set, the threshold for wave breaking of intermediate water waves can be determined as $0.835 \pm 0.005$. Although this is slightly lower than the threshold stated by Saket *et al.* (2017) for two-dimensional deep water wave breaking, the data of Saket *et al.* (2017) conform to this revision. This value is also slightly lower than the value proposed numerically by Barthelemy *et al.* (2015b).

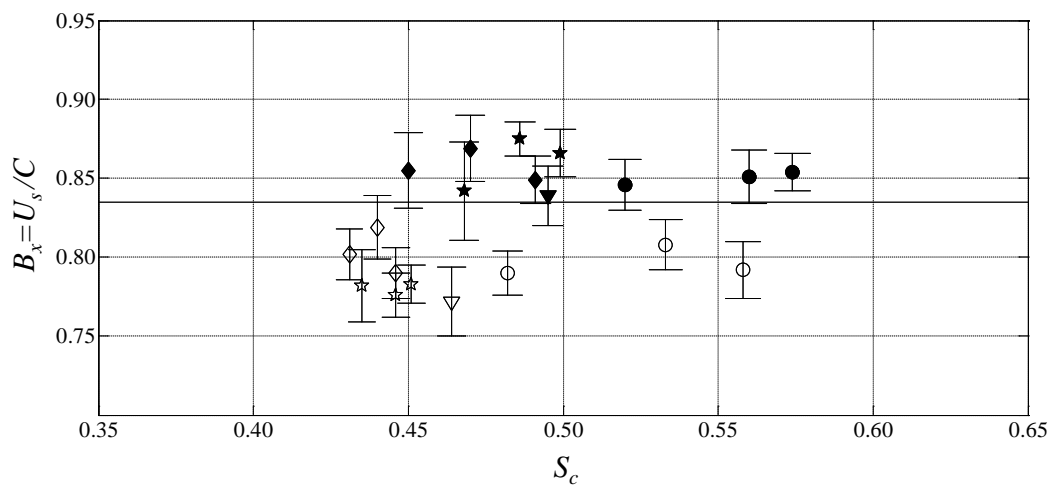

Figure 3. Local wave steepness $S_c$ versus the ratio of water and crest point surface speed $B_x=U_s/C$. C2N3 ($\diamond$), scaled C2N3 ($\triangledown$), C2N5 ($\star$), C3N9 ($\circ$) with maximum recurrence (hollow shapes) and marginal breaking (solid shapes). Error bars indicate the standard deviation of the repeat measurement set (6 to 8 replicates) and the horizontal line at $B_x = 0.835$ is the breaking threshold.

Note that a single C2N3 case was repeated in the same facility but with the spectral frequencies downshifted as much as possible given the available fetch. Both measurements were captured with high precision and obtained with between 6 to 8 independent replicates. The wave steepnesses matched within 1% and the $B_x$ matched within 2%, substantially less than the standard error in each parameter. At the scales achievable in a large-scale laboratory, there is no evidence of threshold dependence on peak spectral wavenumber.

A non-dimensional parameter ($\lambda\, \rho^{0.5} g^{0.5} \sigma^{-0.5}$) was also used to evaluate the dependency of the breaking threshold $B_x$ on the wave scale, where $\rho$ is the density of water, $g$ is the acceleration due to gravity and $\sigma$ is the surface tension. The water density and surface tension were determined assuming pure water at the water temperature measured during each experiment. This dimensionless parameter versus energy flux ratio values for maximum recurrence and marginal breaking waves in intermediate water are shown in figure 4. As seen, no dependency of the breaking threshold on the wavelength can be discerned.

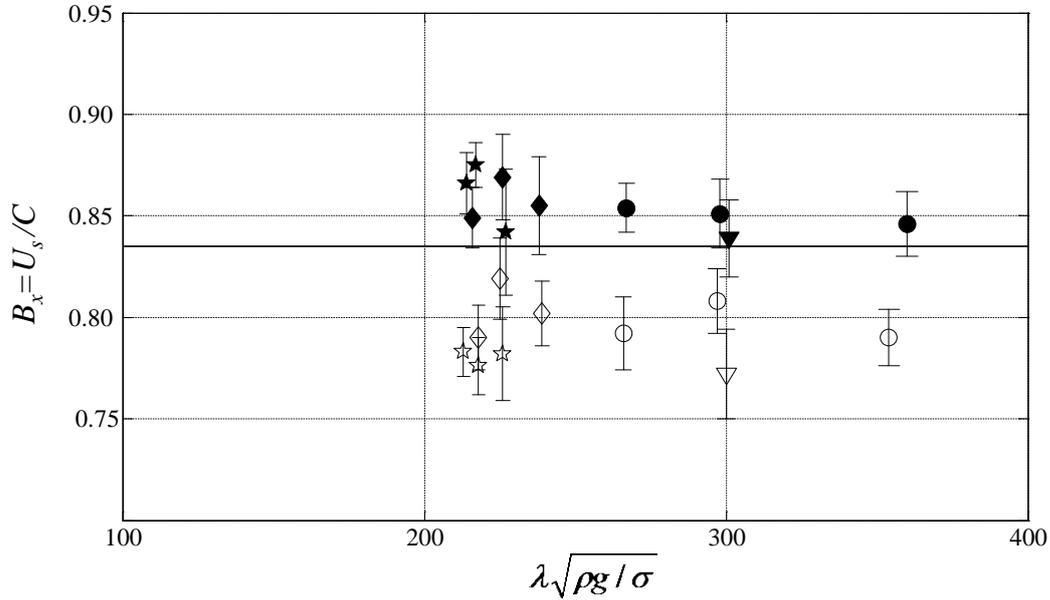

Figure 4. Measured values of $B_x = U_s/C$ as a function of $\lambda\sqrt{\rho g/\sigma}$. The shapes are as defined in figure 3. Error bars indicate the standard deviation of the repeat measurement set (6 to 8 replicates) and the horizontal line at $B_x = 0.835$ is the breaking threshold.

3.2 Depth limited wave heights

The present investigation has defined a threshold between non-breaking and breaking waves. Although this does not impose a limit on breaking wave height, the commencement of breaking does indicate the onset of significant wave dissipation.

The parametric characterisation of the limiting wave heights developed empirically by Nelson (1994) for monochromatic waves on horizontal bed was determined for different types of wave groups in the present study and summarised in table 1 and figure 5. In figure 5 also Nelson's characterisation is presented with data assembled from studies relating to breaking onset. The theoretical limits for intermediate and shallow water depths determined by Miche (1944) Equation (1.1) and linear wave theory are also shown.

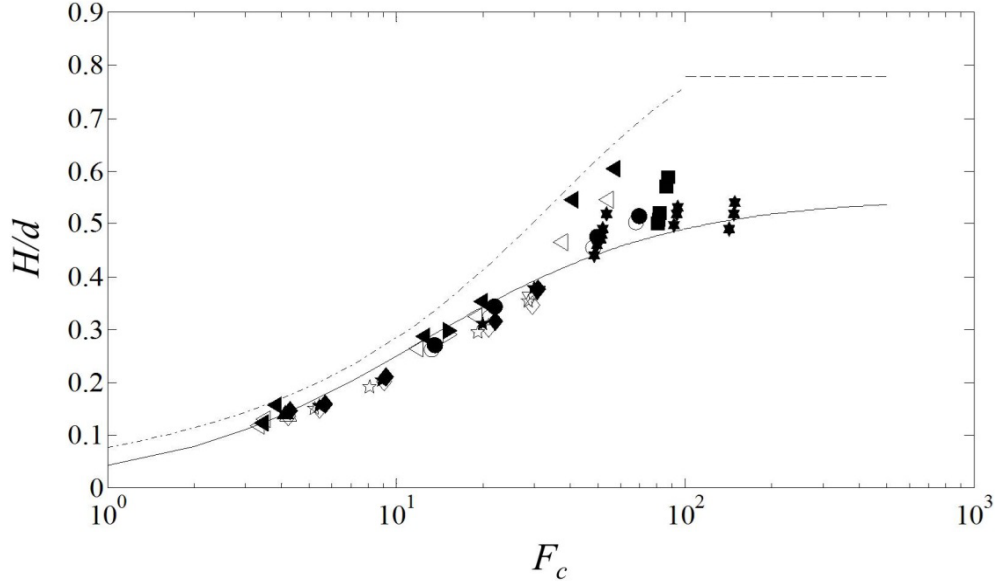

Figure 5. $F_c$ versus $H/d$ for waves on horizontal beds from deep to shallow water for C2N3 (◇), scaled C2N3 (▽), 3D C2N3 (△), C2N5 (☆), C3N9 (○), 3D C3N9 (▷) wave groups in the present study, Barthelemy *et al.* (Class 3 waves, 2013) (◁), Riedel & Byrnes (random waves, 1986) (✿), Dack & Peirson (Class 2 waves, 2005) (□). Maximum recurrence and breaking waves are shown as hollow and solid shapes respectively. Nelson's characterisation (monochromatic, 1994) (———), Miche (1944) theoretical limit (— - —) and McCowan (1894) theoretical limit of 0.78 (— — —).

As shown, up to a breaker index of approximately 0.45, the characteristics of the waves at the threshold of this present investigation track Nelson's curve closely. Marginally breaking waves of this present study remain with 10 % of values specified by Nelson's curve. Consequently, there should be close correspondence between the surface velocities predicted by Nelson's characterisation and the threshold determined here.

The most developed method of computing crest surface velocities are the steady wave methods described by Fenton (1990). Fenton's spectral approach was applied to the Nelson's curve up to $H/d = 0.5$ (excluding values in the vicinity of 0.2 where the method did not converge) and yielded a ratio of $U_s/C$ of $0.49 \pm 0.04$.

While the consistency of this ratio mirrors reasonably the finding of this present study, the mean value of the ratio determined is substantially less. There are two contributing factors. First, as shown by Banner *et al.* (2014) and Barthelemy *et al.* (arXiv:1508.06001v1, 2015a, figure 10), group waves go through a systematic leaning cycle that leads to a slowdown in largest waves within a group. Secondly, as shown by Barthelemy *et al.* (2015a, figure 15), the surface velocities of waves transitioning through a group maximum are substantially in excess of those predicted by higher order steady wave theories.

The findings of a sequence of studies over many years now call into question the limit determined empirically by Nelson (1994). The sequence of studies that are summarised in figure 5 show that for shallower water, the effect of wave grouping can generate extreme waves that are at least 30 % greater than the values determined by Nelson. The McCowan (1894) and Miche (1944) theoretical limits appear conservative and our present studies would support these as an appropriate basis for design with the curves as shown in figure 5. However, it is important to note that this present study has only investigated marginally breaking waves.

More strongly breaking group waves may exceed the McCowan/Miche limits. Allis (2013) systematically investigated more strongly breaking waves in deep water and observed that the crest steepnesses of strongly breaking 3D waves could exceed those of recurrent waves by up to 32%. The mixed nature of the axes of figure 5 means that changes in water height will lead to equivalent shifts in both axes directions. In deeper water, strong shifts from the trajectory of the Nelson curve are not anticipated. However, similar shifts in strongly breaking 3D waves in shallower water would yield values significantly higher than those in the presently assembled data in figure 5. This requires further investigation.

## 4. Conclusions and recommendations

Large-scale laboratory experiments have been conducted for different classes of unforced unidirectional wave groups in intermediate water depths to investigate the breaking threshold proposed by Barthelemy *et al.* (2015b). The measurement techniques used were identical to those used by Saket *et al.* (2017).

The threshold of breaking which distinguishes maximal recurrent waves from marginally breaking waves was found to be $B_x = 0.835 \pm 0.005$. This result is also robust for the deep water experimental results of Saket *et al.* (2017). It is slightly less than the threshold proposed by Barthelemy *et al.* (2015b).

The sensitivity of the threshold in relation to wave scale was investigated and no systematic changes in the threshold with scale could be discerned.

The breaker indices of the waves of the present study were compared with the limiting wave heights determined by Nelson (1994). Nelson's characterisation gives a reasonable estimation of the breaker index to approximately 0.45 ($F_c<40$) but for higher $F_c$ the effect of wave grouping can generate extreme waves that are significantly larger than the limiting wave heights determined by Nelson (1994) for shallow water.

**Acknowledgements**

Funding for this project was provided by the Australian Research Council under Discovery Project DP120101701. Technical assistance provided by Messrs. Larry Paice and Robert Jenkins is gratefully acknowledged.